\documentclass[twocolumn,superscriptaddress,prl]{revtex4-2}

 \usepackage{tabularx}

 \usepackage{float}

\usepackage{parskip}

\usepackage{amsthm}

\usepackage{stix}

 \usepackage{amsfonts}
\usepackage{amsmath}
\usepackage{amssymb}
\usepackage{graphicx}
\usepackage{dcolumn}
\usepackage{dsfont}
\usepackage{times}
\usepackage{epsfig}
\usepackage[]{units}

\usepackage{lipsum}

\newcommand\R{\mathbb{R}}

\newcommand\sech{\mathrm{sech}}

\renewcommand{\arraystretch}{1.5}

\begin{document}

\title{\textbf{Modulation Instability and Wavenumber Bandgap Breathers in a Time Layered Phononic Lattice}}

\author{Christopher Chong}
\affiliation{Department of Mathematics, Bowdoin College, Brunswick, Maine 04011}

\author{Brian Kim}
\affiliation{Department of Mechanical and Civil Engineering, California Institute of Technology, Pasadena, CA 91125, USA.}

\author{Evelyn Wallace}
\affiliation{Department of Mathematics, Bowdoin College, Brunswick, Maine 04011}

\author{Chiara Daraio}
\affiliation{Department of Mechanical and Civil Engineering, California Institute of Technology, Pasadena, CA 91125, USA.}

\begin{abstract}
We demonstrate the existence of wavenumber bandgap (q-gap) breathers in a time-periodic
phononic lattice. These breathers are localized in time and periodic in space,
and are the counterparts to the classical breathers found in space-periodic systems.
We derive an exact condition for modulation instability that leads to the opening of wavenumber bandgaps.
The q-gap breathers become more narrow and larger in
amplitude as the wavenumber goes further into the bandgap. In the presence of damping, these structures acquire
a non-zero, oscillating tail. The experiment and model exhibit qualitative agreement.
\end{abstract}

\date{\today}

\vspace{1cm}

\maketitle

% %%%%%%%%%%%%%%%%%%%%%%%%%%%%%%%%%%%%%%%%%%%%%%%%%%%%%%%%%%%%%%%%%%%%%%%%%%%%%%

\section{Introduction}
\vspace{-.25cm}
The classical discrete breather is defined as a spatially localized, time-periodic solution of a nonlinear lattice differential equation. They are a fundamental structure found in many platforms, including photonics, phononics,
and in electrical systems, see \cite{flach_discrete_2008,pgk:2011,Dmitriev_2016}
for comprehensive reviews. One mechanism through which breathers can manifest is the
modulation instability (MI) of plane waves in spatially
periodic lattices~\cite{Huang2}.
Such breathers have a frequency that falls into a spectral gap~\cite{flach_discrete_2008}.

A natural counterpart to the fundamentally important discrete breather is
the so-called wavenumber bandgap (q-gap) breather. It 
is localized in time, periodic in space, and has
wavenumber that falls into a q-gap. 
Q-gap breathers represent a newer class of solutions
and are distinct from q-breathers, which are localized in wavenumber
and periodic in time \cite{Qbreather}.
Since q-gap breathers
have the roles of time and space switched when compared to classical breathers,
it is natural to consider lattices that
are time varying (instead of space varying).
Recent advances in experimental platforms for time varying systems
makes the topic particularly relevant, including studies in
photonic \cite{soljacic_optimal_2002,wang_optical_2008,wen_tunable_2020,Rechtsman2021},
electric \cite{powell_multistability_2008,kozyrev_parametric_2006,powell_asymmetric_2009}, 
and phononic systems \cite{cassedy_dispersion_1967, reyes-ayona_observation_2015, wang_observation_2018, galiffi_broadband_2019, trainiti_time-periodic_2019, lee_parametric_2021,trainiti_non-reciprocal_2016,wang_observation_2018,goldsberry_non-reciprocal_2019,chen_nonreciprocal_2019,zhu_non-reciprocal_2020,marconi_experimental_2020,nassar_nonreciprocity_2020, Kim2023}.
Time localization can also arise via mechanisms other than a wavenumber bandgap,
such as zero-wavenumber gain modulation instability \cite{ZeroGain}.
The so-called Akhmediev breathers of
the nonlinear Schr\"odinger equation \cite{AkhmedievNLS}, and its discrete integrable counter-part, the Ablowtiz-Ladik
lattice \cite{AkhmedievAL}, are also localized in time. They do not have wavenumber in a q-gap,
and hence are distinct from q-gap breathers. 
Very recently, it was shown theoretically that time-localized solitons with wavenumbers in the q-gap exist in a time-periodic photonic crystal \cite{Superluminal}.
Breathers in time varying phononic systems, however, remain unexplored. 

In the present letter, we employ numerical, analytical, and experimental approaches to 
explore q-gap breathers and the modulation
instability leading to the opening of wavenumber bandgaps. Q-gap breathers could be exploited
for the creation of phononic frequency combs \cite{FrequencyComb1,FrequencyComb2,MicroBreather}, energy
harvesting applications \cite{harvesting2,harvesting3}, or acoustic signal processing \cite{Hartmann2007}.

%%%%%%%%%%%%%%%%%%%%%%%%%%%%%%%%%%%%%%%%%%%%%%%%%%%%%%%%%%%%%%%%%%%%%%%%%%%%%%
\vspace{-.5cm}
\section{Experimental and Model Set-up} \label{sec:exp}
\begin{figure}
 \includegraphics[height=2.5cm]{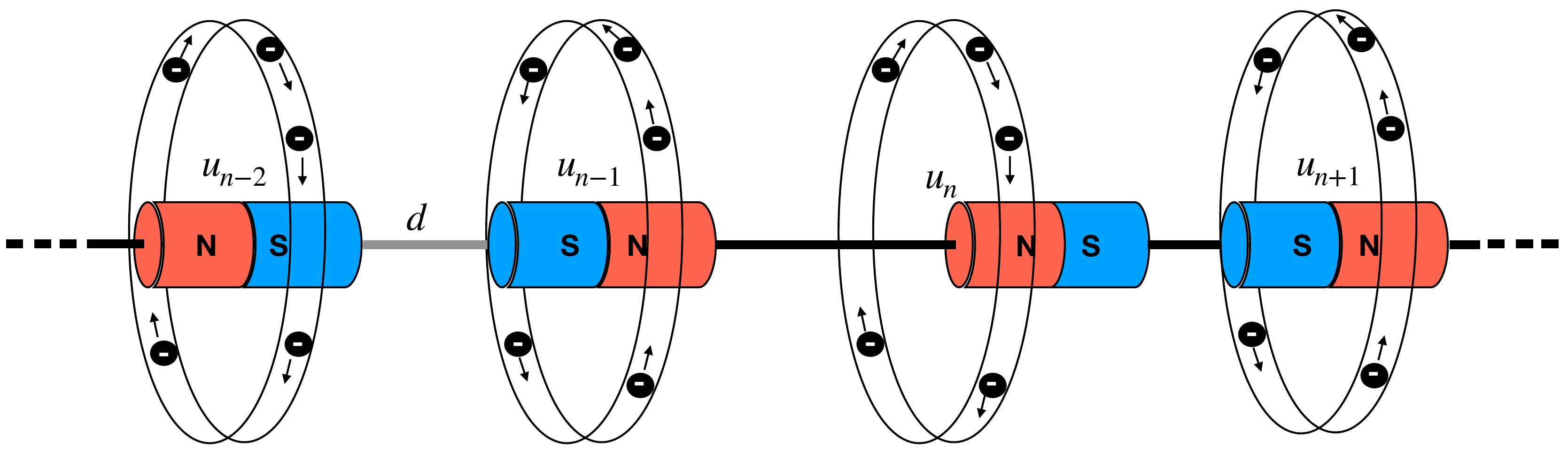} 
    \caption{Schematic of the modulated magnetic lattice.}
    \label{fig:schematic_photo}
\end{figure}

\vspace{-.25cm}
The particular phononic lattice under consideration is a system of
repelling magnetic masses with grounding stiffness controlled by electrical coils driven applied voltage signals that are given by a time-periodic step function. The experimental setup is adapted form the platform developed in \cite{wang_observation_2018,Kim2023}. The chain is composed of $N-1$ ring magnets (K\&J Magnetic, Inc., P/N R848) lined with sleeve bearings (McMaster-Carr P/N 6377K2) comprising the uniform masses, arranged with alternating polarity on a smooth rod (McMaster-Carr P/N 8543K28). Electromagnetic coils (APW Company SKU: FC-6489) are fixed concentrically around the equilibrium positions of each of the innermost eight masses, such that they may exert a restoring force on each mass proportional to the current induced by applied voltage step-function (Aglient 33220A, Accel Instruments TS250-2).
%The chain has fixed boundary conditions.
%and the input mass (the first free inward mass from one of the fixed ends) has a concentric electromagnetic coil offset axially from its equilibrium position, which provides the driving force. 
The velocity of each mass is measured using laser Doppler vibrometer (Polytec CLV-2534), repeating experiments to a construct full velocity field for the lattice.
Figure \ref{fig:schematic_photo} shows a schematic of the experimental setup. 
The system is modeled as Fermi-Pasta-Ulam-Tsingou type lattice  \cite{wang_observation_2018,Kim2023}
%
% \begin{align} \label{eq:EOM} 
% 	M  \ddot{u}_n    &   +   k(t) \, u_n   +  c \dot{u}_n =  \\
%  &A \left( d + u_n - u_{n-1} \right)^{-\alpha}  
% 	 - A \left(d + u_{n+1}  - u_n \right)^{-\alpha}	 \nonumber
% \end{align}
\begin{align} \label{eq:EOM} 
	M  \ddot{u}_n      +   k(t) \, u_n   +  c \dot{u}_n =  F(u_n - u_{n-1}) - F(u_{n+1}  - u_n)
\end{align}
where $u_n$ is the displacement of the $n$th ring magnet from its equilibrium position,
where the equilibrium distance between adjacent magnets is $d=0.0334$ m. The indices run from $n=1 \ldots N-1$ and we consider fixed boundary conditions $u_{0}(t) = u_{N}(t)=0$. All ring magnets have uniform mass $M=0.0097$ kg. Dissipative forces are modeled with a phenomenological
viscous damping term  $c \frac{d {u}_{n}}{d t}$,
where the damping coefficient $c=0.15$ Ns/m is determined empirically by matching the simulated and experimental spatial decay of the velocity amplitude envelope of waves traveling through the lattice \cite{Kim2023}. The coupling force term is defined using the repulsive magnetic force between neighboring masses. The experimentally measured force-distance relation between neighboring masses is fit with a dipole-dipole approximation, as in \cite{wang_observation_2018}, which is given by $F(x) = A(d+x)^{-\alpha}$, where $x$ is the center-to-center distance between masses with $A = 9.044\times10^{-7} \text{Nm$^{4}$} $ and $\alpha = 4$. The current resulting from a periodic step function voltage applied to the electromagnetic coils induces a magnetic field that provides a grounding stiffness modulation of the form
\begin{equation}\label{eq:modulation}
	k(t)= \left \{
\begin{array}{cc}
	k_a  &  0 \leq t < \tau T \\
	k_b &  \tau T \leq t < T \
\end{array} \right. \,, \quad  k(t) = k(t + T) %for all $t$.
\end{equation}
%where we assume that $k(t)$ is periodic with period $T$,
%namely,  %see Fig.~\ref{fig:schematic_photo}(b).
The step values $k_a,k_b$ and duty-cycle
$0 < \tau < 1$  ($[\tau] = \text{s}$) are parameters.
Unless otherwise stated, we use $k_a = 0$, $k_b = 150$ N/m and $\tau = 0.5$ s.
We will use the modulation frequency $f_{\rm mod} = 1/T$ Hz
as the main system parameter to be varied.
\vspace{-.25cm}
\section{Modulation Instability}
\vspace{-.25cm}
To find breathers,  we first need to determine the wavenumber bandgap.
This is achieved by computing the stability of plane waves (i.e., the modulation stability) of the linearized model
\begin{equation}\label{eq:modulated_mass_spring}
  M\ddot{u}_n= K (u_{n-1}-2u_n+u_{n+1})- k(t)u_n - c \dot{u}_n,
\end{equation}
where $K = \alpha A d^{\alpha-1}$. 
%% This could be added back
%The linearized equation is obtained by keeping the linear terms in the Taylor expansion of the nonlinear coupling force $A(d + y)^{-\alpha}$,
%where it is assumed that $|u_n - u_{n-1}| \ll d $ for $n=1 \dots N-1$. 
For time-independent stiffness
$(k(t)=0)$ the undamped ($c=0$) linear equation has the dispersion relationship $\omega_{\rm disp}^2(q) = 4 K / M \sin^2(q/2)$
such that the linear spectrum extends from $[0, \sqrt{K/M}]$ and all plane waves are stable.
\begin{figure}
\centerline{
   \begin{tabular}{@{}p{0.5\linewidth}@{}p{0.5\linewidth}@{}  }
  \rlap{\hspace*{5pt}\raisebox{\dimexpr\ht1-.1\baselineskip}{\bf (a)}}
 \includegraphics[height=3.5cm]{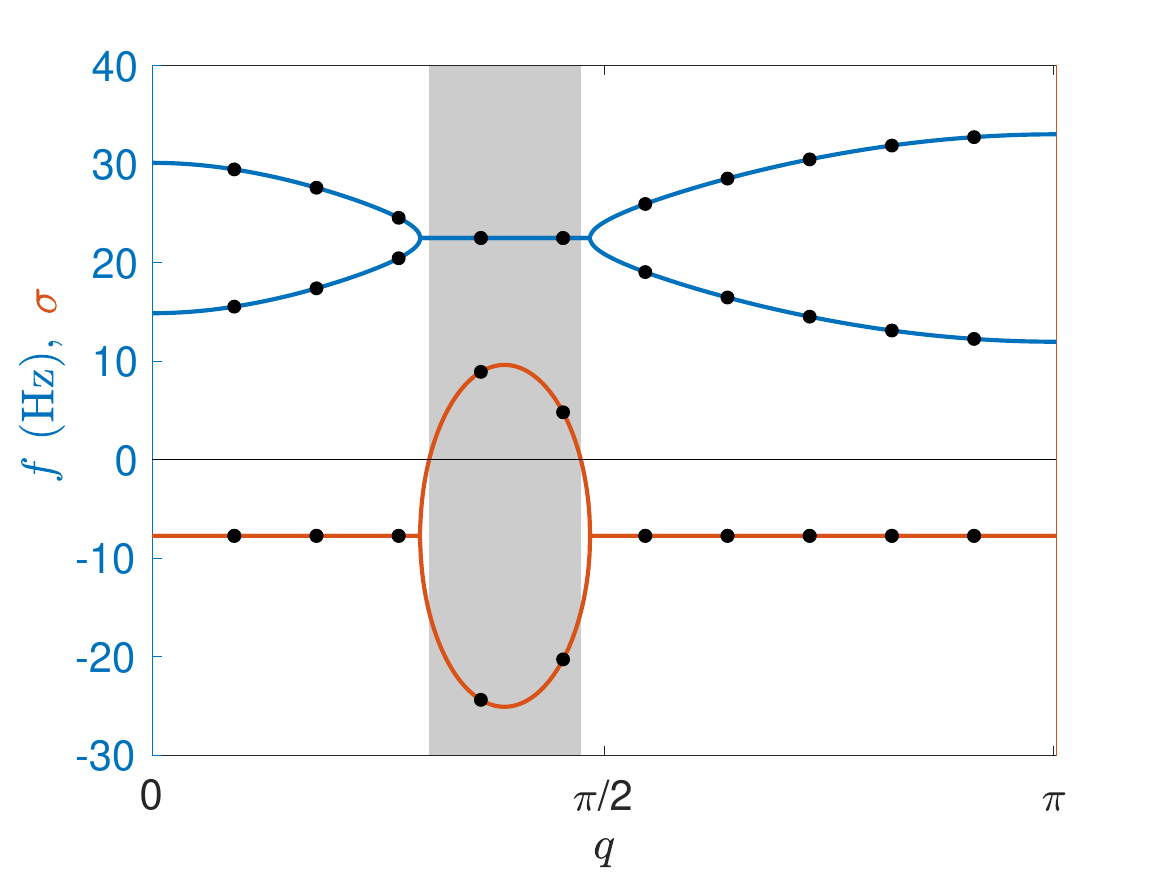} &
   \rlap{\hspace*{5pt}\raisebox{\dimexpr\ht1-.1\baselineskip}{\bf (b)}}
\includegraphics[height=3.5cm]{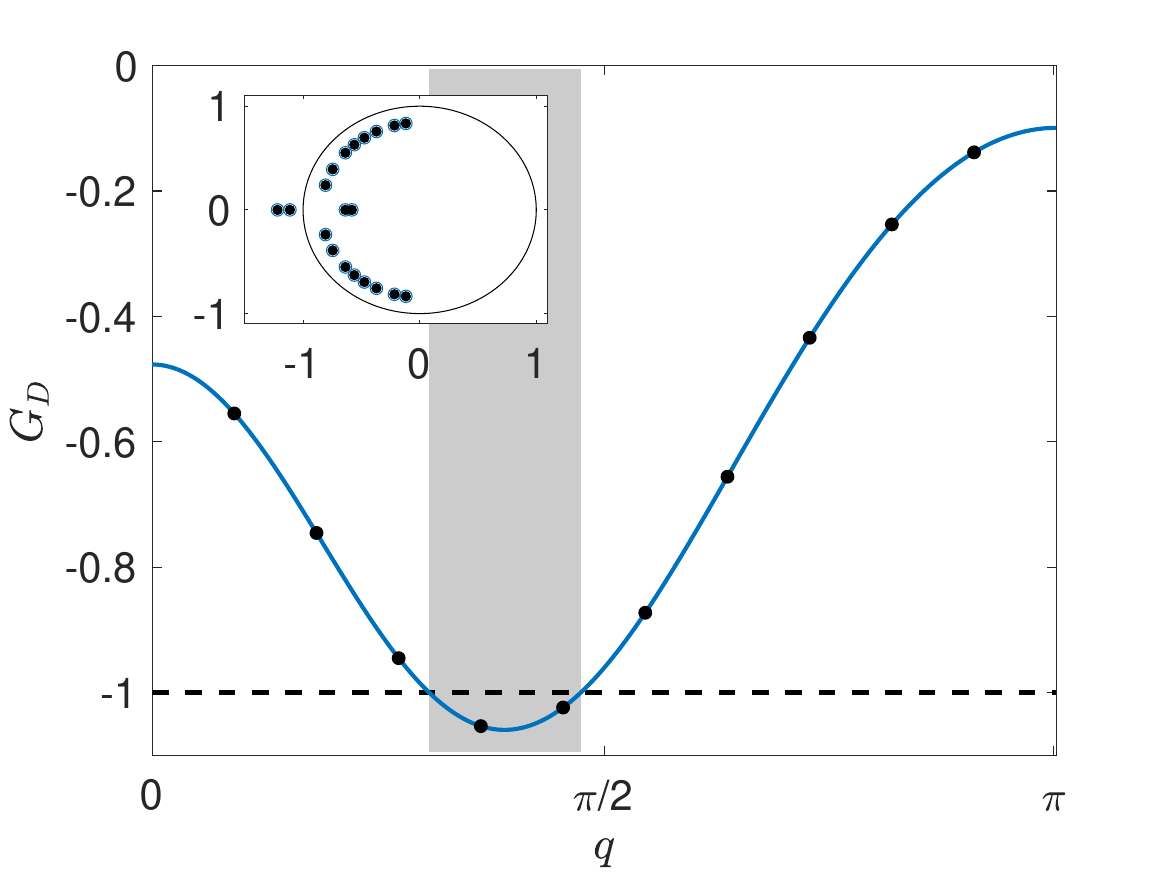}
  \end{tabular}
  }
    \caption{ \textbf{(a)} Plot of $f = \omega / ( 2\pi)$ where $\omega$ is the imaginary part
    of the Floquet exponent (blue curve) and $\sigma$, the real part of the exponent (red curve) for an infinite lattice. The black dots show the corresponding
    values for a finite sized lattice with $N=11$. The modulation frequency is $f_{\rm mod} = 1/T = 45$ Hz.
    %The parameters used are $k_a = 0$, $k_a = 150$, $T = 1 / 45 $, $\kappa = 0.5$ and $c = 0.15$.
    The shaded gray region indicates the region of instability (i.e., the wavenumber bandgap).
    \textbf{(b)}  Plot of the function $G_D$ from Eq.~\eqref{eq:damped_stability} in the infinite lattice
    (solid blue line) and in the $N=11$ lattice (solid black dots). %The modulation frequency is $f_{ \rm mod} = 1/T = 45$ Hz.
    The plane wave is unstable when $|G_D| > 1$, which is highlighted by the gray region.
    The inset shows the corresponding Floquet multipliers in the complex plane. There are two multipliers lying outside the unit circle (also shown)
    demonstrating the instability of a general solution. 
    } 
    \label{fig:Gdisp}
\end{figure}
In the case of time-dependent stiffness $(k(t)\neq 0)$ a gap in the wavenumber axis $q$ is possible.
For general time-periodic stiffness $k(t)$ with period $T = 2 \pi / \omega_{\rm mod}$, a wavenumber bandgap will open
where the dispersion curve $\omega_{\rm disp}(q)$ intersects itself when translated by an integer
multiple of half the modulation frequency $\omega_{\rm mod}/2$ \cite{nassar_modulated_2017}. 
The advantage of considering $k(t)$ to be  a periodic step function is that
the modified dispersion relation can be computed exactly. Making the ansatz $u_n(t)=X_m(n)\cdot \Theta_m(t)$, 
one finds upon substitution into Eq.~\eqref{eq:modulated_mass_spring}
and enforcing Dirichlet boundary conditions that the
eigenfunctions are $X_m(n) = \sin\left( q_m n \right )$
where the wavenumber is $q_m = m \pi/N$
with $m=1,\ldots, N-1$. In the infinite lattice, $q\in[0,\pi]$.
The associated eigenvalues are $\lambda_m = \sin^2\left ( q_m / 2 \right)$.
The temporal part $\Theta(t)$ satisfies
\begin{eqnarray}
%\lambda_m X_m(n) &=& -X_m(n-1) +2X_m(n) - X_m(n+1) \label{eq:Xode}\\
%\ddot{\Theta}_m&=&\frac{-\lambda_m K-k(t)}{M}\Theta_m - \frac{c}{M} %\dot{\Theta}_m   \label{eq:Tode}
M \ddot{\Theta}_m&=& -(\lambda_m K+k(t) )\Theta_m - c \dot{\Theta}_m.   \label{eq:Tode}
\end{eqnarray}
We can obtain an exact solution of this equation (and  
dispersion relation and stability condition), by adapting
a procedure carried out in the context of an undamped Kronig-Penney photonic lattice \cite{Centurion,Rapti_2004}.
The general
solution of Eq.~\eqref{eq:Tode} will be a superposition of functions of the form
$\Theta_m(t) = H_{m}(t) e^{\mu_m t}$ where $H(t)$ has period $T$ and 
%\begin{equation} \label{eq:theta}
%    \Theta_m(t) = H_{m}(t) e^{\mu_m t}, \quad H(t) = H(t +T)   
%\end{equation}
%
$\mu_m = \sigma_m + i \omega_m$ is the Floquet exponent where $\sigma_m , \omega_m \in \R$. 
The Floquet multiplier is $e^{\mu_m T}$. %$H_{m}(t)$ is a function with period $T$.
The waveform
associated to the wavenumber $q_m$ will be stable if $\sigma_m \leq 0$, or equivalently,
if the Floquet multiplier has modulus less than unity, $|e^{\mu_m T}| \leq 1$.
Substitution of $\Theta_m(t) = H_{m}(t) e^{\mu_m t}$ into equation~\eqref{eq:Tode} and 
demanding that $H_m(t)$ and $\dot{H}_m(t)$ are continuous 
at $t=0$ and $t=\tau$ leads to the following equations (detailed in the Supplemental Material)
\begin{eqnarray}\label{eq:realpart}
    G&=& \cosh\left[ \left(\sigma_m +\frac{c}{2M} \right) T\right] \cos(\omega_m T ) \\
    0&=&\sinh\left[ \left(\sigma_m +\frac{c}{2M} \right) T\right]\sin(\omega_m T ) \label{eq:imagpart} 
\end{eqnarray}
where $G$ depends on the wavenumber and system parameters, but not the Floquet exponent $\mu_m$:
\begin{align}
%G(q,k_a,k_b,\tau,T)
%& G \equiv \\&
G \equiv &- \frac{s(k_a)^2+s(k_b)^2}{2s(k_a)s(k_b)}        \sin(s(k_a)\tau T)\sin(s(k_b) (1-\tau) T)  \nonumber \\
&+ \cos(s(k_a)\tau T)\cos(s(k_b) (1-\tau ) T)   \nonumber 
\end{align}
where
$s(k)=\sqrt{4M\left(\lambda_m \cdot K+ k  \right) -c^{2}} / \left( 2M \right)$.
 These equations allow for the exact computation of the Floquet exponents
 $\mu_m = \sigma_m + i \omega_m$. An example plot is shown in Fig.~\ref{fig:Gdisp}(a).
 %Given a fixed set of parameters and wavenumber, $q_m$, 
 If $|G| \leq 1$ then $\sigma_m = - c/(2M)$ and
 $\omega_m = \cos^{-1}(G)/T$. In this case the underlying solution is stable.
If $ \pm G > 1$ then $\omega_m  = (3 \pm 1) \pi /(2 T)$,
implying that the imaginary part of the Floquet exponent is an integer multiple of half the modulation frequency.
In this case, the real part
of the Floquet exponent is $\sigma_m = \pm \cosh^{-1}\left( \mp G  \right)/T - c/(2M)$
which implies the following condition for stability,
 \begin{equation} \label{eq:damped_stability}
 \left | G_D \right | \leq 1, \qquad G_D \equiv G \, \mathrm{sech}\left( \frac{c T}{2M} \right).
 %\left |\left( G\pm \sqrt{G^2-1}\right)e^{-\frac{c T}{2M}    }  \right|>1
 \end{equation}
Note that this expression is exact and gives an 
efficient way to check for stability via
direct substitution of the system parameters into $G_D$ and simply checking the inequality. 
A plot of $G_D$ is shown in Fig.~\ref{fig:Gdisp}(b).
The linear stability predictions agree well with experimental observations
(examples given in the Supplemental Material).
The set of wavenumbers where $|G_D| >1$ make up the so-called wavenumber bandgap.
The edges can be found by solving $G_D = \pm 1$. See
the gray regions of Fig.~\ref{fig:Gdisp} for example wavenumber bandgaps.

\begin{figure*}
\centerline{
   \begin{tabular}{@{}p{0.33\linewidth}@{}p{0.33\linewidth}@{}p{0.33\linewidth}@{} }
     \rlap{\hspace*{5pt}\raisebox{\dimexpr\ht1-.1\baselineskip}{\bf (a)}}
 \includegraphics[height=4.7cm]{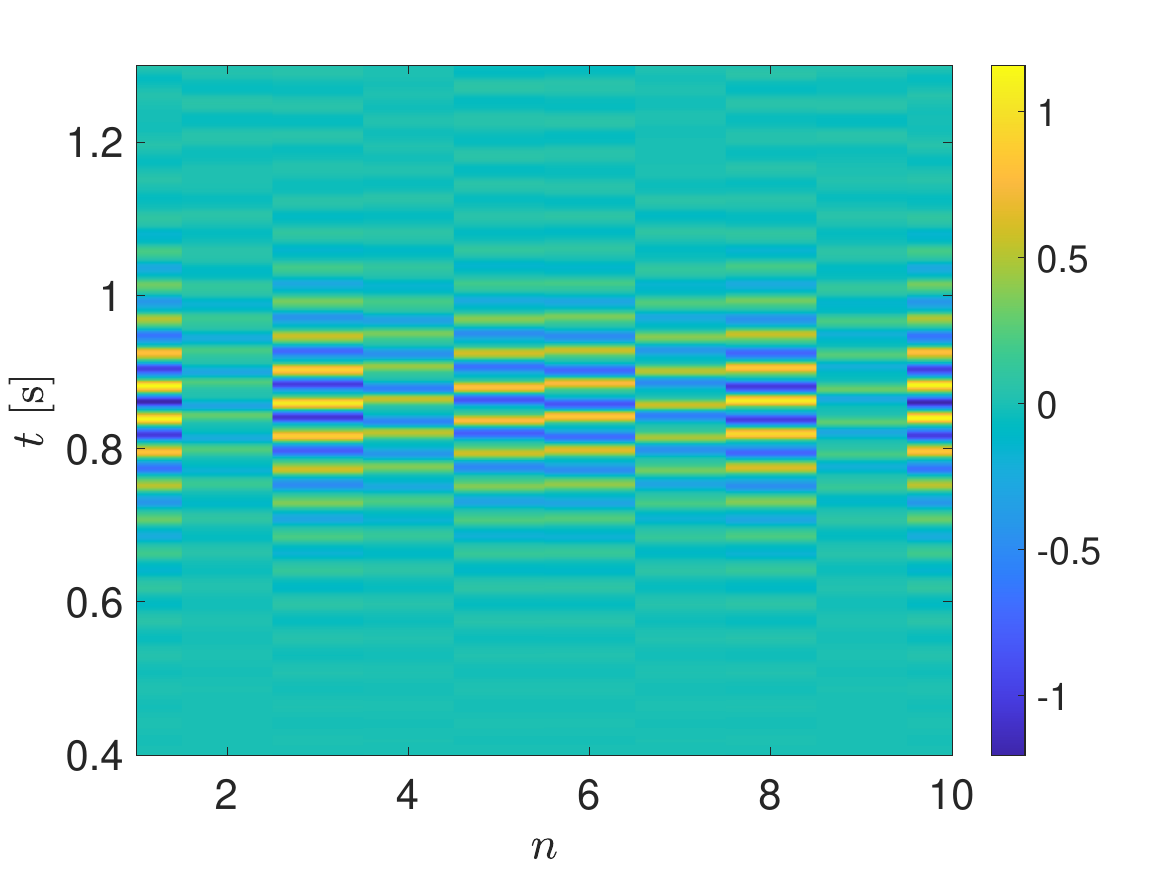} &
  \rlap{\hspace*{5pt}\raisebox{\dimexpr\ht1-.1\baselineskip}{\bf (b)}}
 \includegraphics[height=4.7cm]{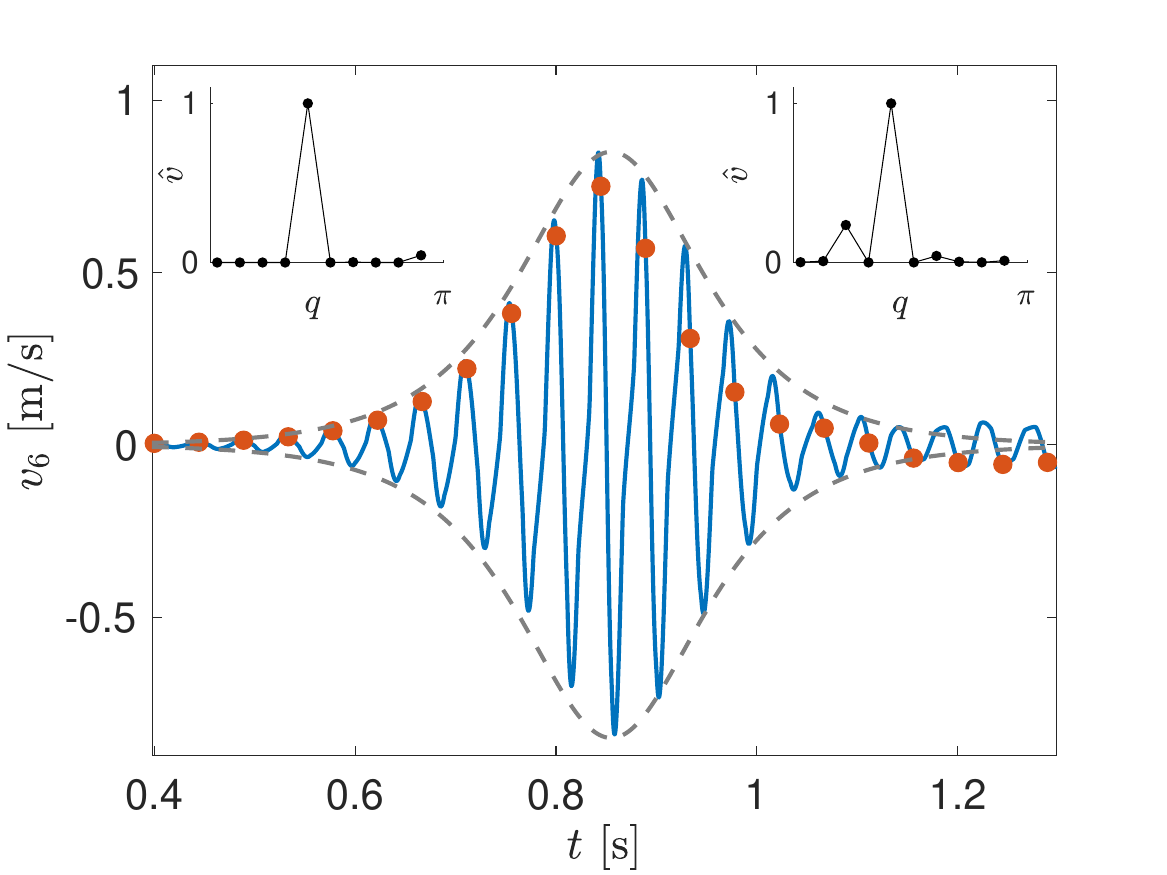} &
   \rlap{\hspace*{5pt}\raisebox{\dimexpr\ht1-.1\baselineskip}{\bf (c)}}
 \includegraphics[height=4.7cm]{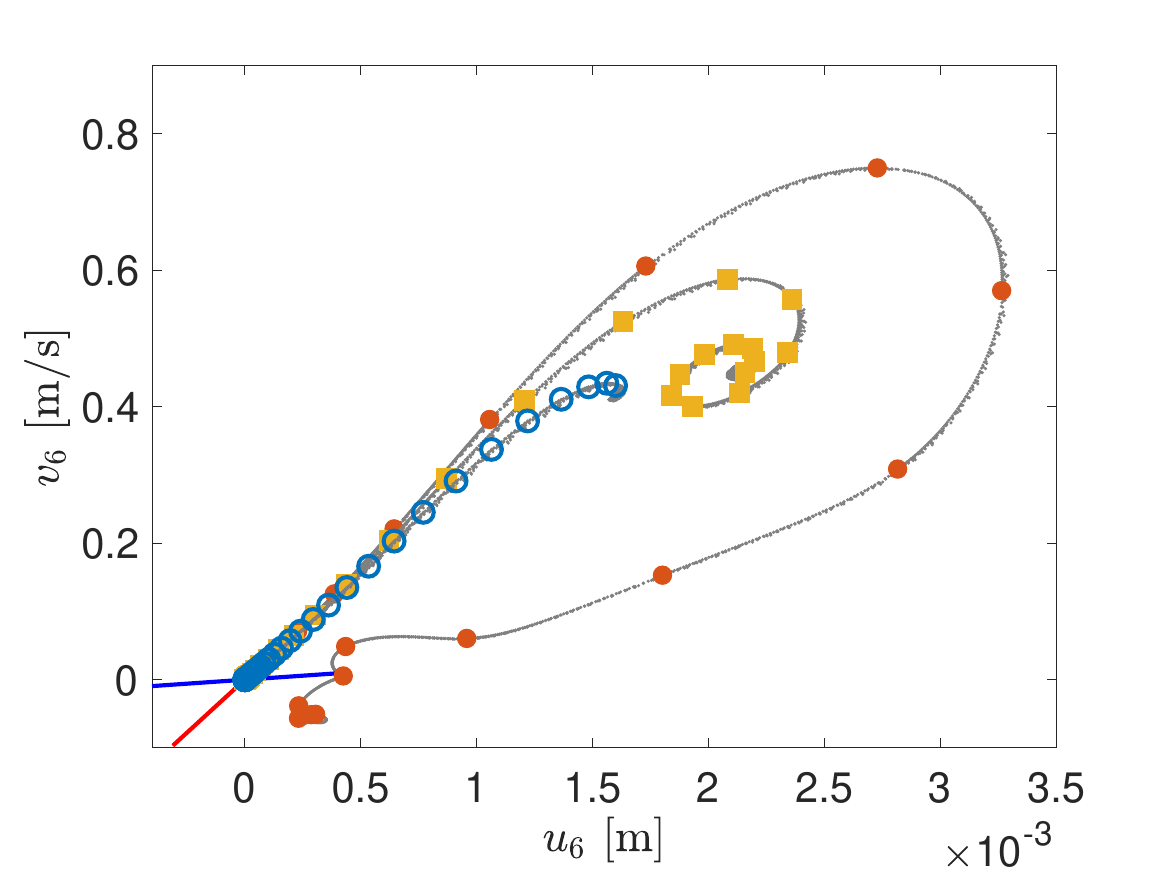} 
  \end{tabular}
  }
    \caption{Wavenumber bandgap breather for $f_{\rm mod} = 45$ Hz. \textbf{(a)} Intensity plot of velocity after initializing the unstable plane wave
    with wavenumber $q_5$ with $c=0$. \textbf{(b)} Time series of the velocity of node 6 of panel (a). The dashed line
    shows the best-fit envelope (in the least square sense). The red dots are the solution sampled every $2T$ seconds. The insets show
    the spatial Fourier transform before (left) and after (right) the turning point. \textbf{(c)} Plot
    of the Poincar\'e map of the solution shown as red dots in panel (b). The yellow squares and blue circles
    correspond to the same simulation with $c=0.075$ Ns/m and $c = 0.15$ Ns/m, respectively. 
   } 
    \label{fig:breather_ns10}
\end{figure*}

\vspace{-.5cm}
\section{Wavenumber Bandgap Breathers}
\vspace{-.25cm}
To draw analogy to the classic breathers of space-periodic systems, 
we first consider Eq.~\eqref{eq:EOM} without damping $(c=0)$. To generate a wavenumber bandgap breather, Eq.~\eqref{eq:EOM} is
%is simulated with an initial value given by an
initialized with an unstable plane wave, e.g., $u_n(0) = a \sin\left( q_m n\right)$,
where $q_m \in (q_\ell,q_r)$ where $(q_\ell,q_r)$ is the wavenumber bandgap, $0<a \ll 1$, and the initial velocity $v_n(0)$ is determined via Eq.~\eqref{eq:Tode}.
An example is shown in
Fig.~\ref{fig:breather_ns10} with $f_{\rm mod}=45$ Hz,  $a=10^{-4}$ %where the system parameters are the same as in Fig.~\ref{fig:schematic_photo}(c)
and wavenumber $q_5= 5 \pi / 11$, which falls 
into the gap $(q_\ell,q_r) \approx   (0.93,1.52)$. %(0.899,1.441)$. 
%The wavenumber is quite close to right bandedge. In particular
%the distance to the edge is $\Delta q  = q_r - q_5 \approx 0.09$.
The Floquet exponent associated to this wavenumber is $\mu_5 = \sigma_5 + i \omega_5 = 4.8107 +  i \omega_{\rm mod}/2 $.
The corresponding
Floquet multiplier is shown in the inset of Fig.~\ref{fig:Gdisp}(b) (the smaller of the two multipliers
lying outside the unit circle).

\begin{figure*}[t]
\centerline{
   \begin{tabular}{@{}p{0.33\linewidth}@{}p{0.33\linewidth}@{}p{0.33\linewidth}@{} }
  \rlap{\hspace*{5pt}\raisebox{\dimexpr\ht1-.1\baselineskip}{\bf (a)}}
 \includegraphics[height=4.7cm]{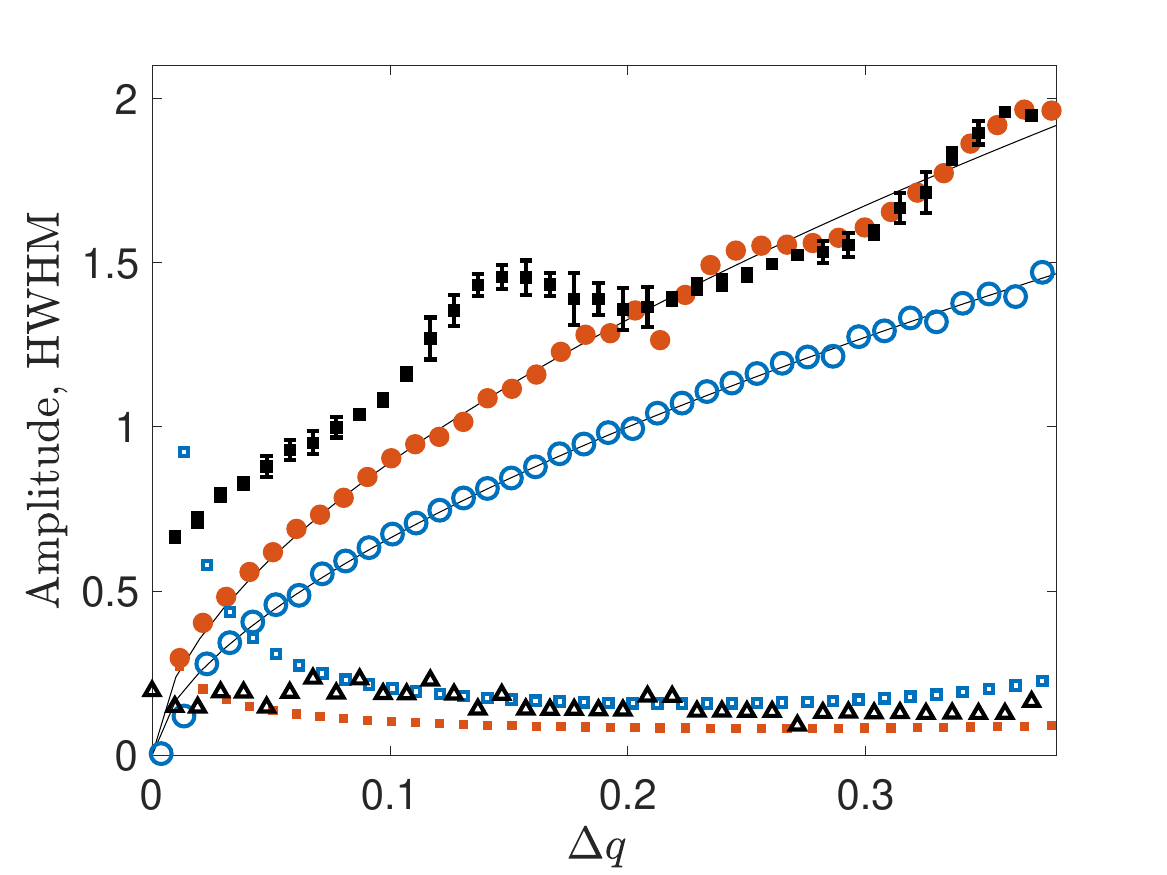} &
   \rlap{\hspace*{5pt}\raisebox{\dimexpr\ht1-.1\baselineskip}{\bf (b)}}
 \includegraphics[height=4.7cm]{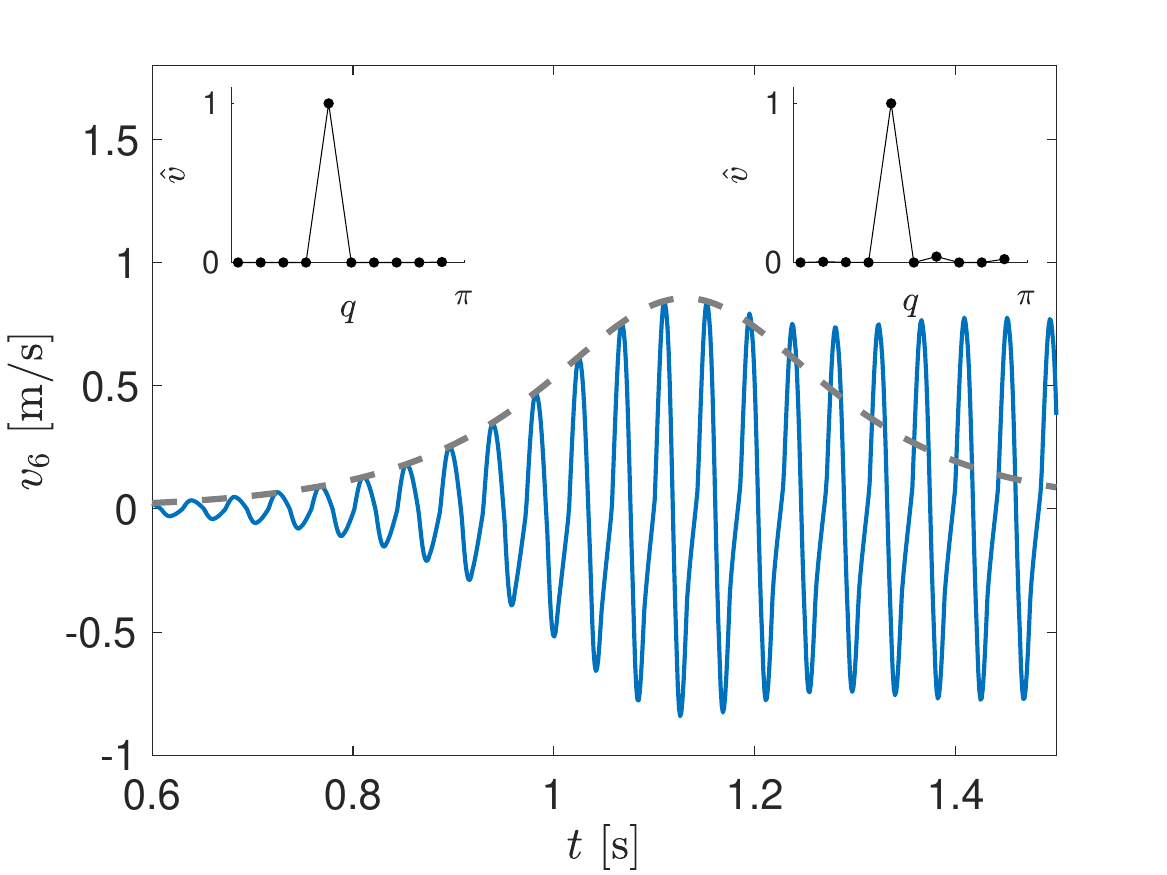} & 
    \rlap{\hspace*{5pt}\raisebox{\dimexpr\ht1-.1\baselineskip}{\bf (c)}}
 \includegraphics[height=4.7cm]{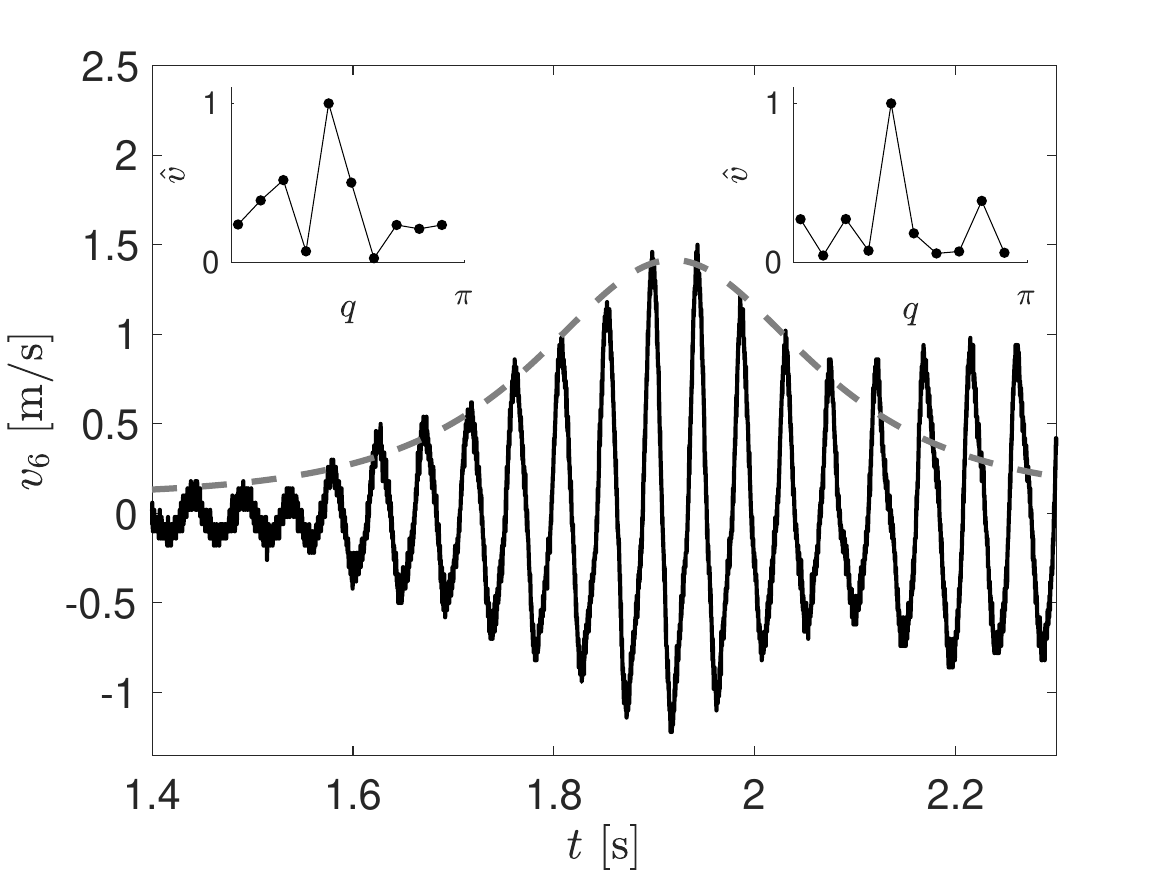} 
  \end{tabular}
  }
    \caption{\textbf{(a)} Plot of the breather amplitude for the $c=0$ (red dots) and $c=0.15$ Ns/m (blue circles) simulations vs.
    distance of the wavenumber to the bandedge, $\Delta q$. The lines
    show the best fit function of the form $\beta_1 \Delta q ^{\beta_2}$. The markers with error bars are the measured amplitude from the experiment. The solid red squares and open blue squares show the theoretical HWHM for the $c=0$ and $c=0.15$ Ns/m cases, respectively. The black
    triangles are the experimentally measured HWHM. \textbf{(b)} Time series example for the $c=0.15$ Ns/m simulation
    with $\Delta q = 0.147$.  The dashed line
    shows the best-fit envelope of the solution up until the maximum. The insets show
    the spatial Fourier transform before (left) and after (right) the turning point.
    \textbf{(c)} Same as panel (b), but for the experiment.
    } 
    \label{fig:cont}
\end{figure*}

The solution
initially grows exponentially with growth rate given by $\sigma_5$, due to the modulation instability,
but reaches a turning point and then decays with the rate $-\sigma_5$.
This happens
uniformly within the lattice, as shown by the intensity plot in Fig.~\ref{fig:breather_ns10}(a).
Figure~\ref{fig:breather_ns10}(b) shows the time series of the velocity of the 6th node, i.e., $\dot{u}_6(t) = v_6(t)$
(solid blue curve). Both panels (a) and (b) demonstrate that the dynamics are localized in time. Spatial periodicity of the solution is imposed by construction
due to the finite length of the lattice with zero boundary conditions. The role of space and time have been switched when compared to
the classic breathers of space-periodic systems. Thus, the
solution shown in Fig.~\ref{fig:breather_ns10} is the so-called called wavenumber bandgap breather.
% this could go into SM
Motivated by the fact that the envelope of a breather of a space-periodic FPUT lattice is described
by a soliton of the Nonlinear Schr\"odinger (NLS) equation (in the limit
of the temporal frequency approaching the band edge from within the spectral gap),
we fit the velocity profile $v_6$ with a function of the form $\beta_1 \sech\left( \sigma_5 (t - \beta_2) \right)$ where
$\beta_j$ are fitting parameters and $\sigma_5$ is the real part of the Floquet exponent.
See the gray dashed line of Fig.~\ref{fig:breather_ns10}(b). The good agreement
between the velocity profile and the fit envelope function confirms that the growth/decay
rate is indeed given by the real part of the associated Floquet exponent, in this case $\sigma_5$.
To better understand the mechanism behind the formation of the wavenumber bandgap breather,
we construct a Poincar\'e map of the dynamics by sampling the solution with the frequency
associated with the breather, namely $\omega_{\rm mod}/2$. This corresponds to twice
the period of the modulation. Thus, the map will be of the form
 $\mathbf{F}_j(\mathbf{u}^0) = \mathbf{u}(2 T j)$, where $j$ is an integer, $T$ is the period of $k(t)$, and
 $\mathbf{u} = (u_1,u_2,\ldots,u_{N-1},v_1,v_2,\ldots,v_{N-1} )  \in\R^{2(N-1)}$ is vector valued solution of Eq.~\eqref{eq:EOM} with initial value $\mathbf{u}^0$.
As in the simulation shown in Fig.~\ref{fig:breather_ns10}(b), the initial value of the map is given by an unstable plane wave (e.g.,
  with wavenumber $q_5$). The red dots of Fig.~\ref{fig:breather_ns10}(b) show values of 
  the map $\mathbf{F}$ that correspond to $v_6$. %(i.e. the $N-1+6$ component of $\mathbf{F}$).
The red dots of Fig.~\ref{fig:breather_ns10}(c) show values of $\mathbf{F}$
in the $(u_6,v_6)$ phase plane. The eigenvector corresponding to the unstable (stable) Floquet 
exponent $\mu_5$ ($-\mu_5$) is shown in red (blue). The gray line of Fig.~\ref{fig:breather_ns10}(c)
is obtained by repeatedly generating the map $\mathbf{F}$ in the $(u_6,v_6)$ plane for various (small)
multiples of the initial value $\mathbf{u}^0$. The origin is a saddle
type fixed point, and the trajectory forms a  near homoclinic orbit, made possible
by the nonlinearity of the system. The orbit is not exactly homoclinic, since it does
not approach the origin via the stable eigenvector as $t \rightarrow \infty$. Indeed,
as can also be inferred from Fig.~\ref{fig:breather_ns10}(b), the solution does not decay
to zero, but rather it experiences small oscillations. This is due to the existence 
of other modes in the system (e.g., ones with associated multipliers lying on the unit circle),
which are excited during the dynamic evolution. The insets of Fig.~\ref{fig:breather_ns10}(b)
show a normalized spatial Fourier transform of the signal
  before (the left inset) and after (the right inset) the maximum velocity is attained. In particular,
 the quantity $| \hat{v} | / |\max \hat{v}|$ is
 shown against the wavenumber, where $\hat{v}(q) = \frac{2}{N-1 }\sum_n v_n(t)\sin(n q ) $
where $t=0.66$s and $t=1.02$s are the times used to compute the transform
before and after the turning point, respectively. Before the turning point the only prominent wavenumber
is the one associated to the initial value (in this case $q_5 = 5 \pi/11$).
After the turning point, there is an additional mode excited
that lies outside the wavenumber bandgap (in this case $q_3 = 3 \pi/11$). 
It is this mode that is primarily responsible for the non-zero oscillations
at the tail of the breather. If one imposes the additional
criterion that a breather must have tails decaying to zero, then strictly speaking, the
structure found here would be a generalized breather, since the 
orbit is not exactly homoclinic. Classic breathers with tails that do not
decay to zero are common in non-integrable systems,
and have sometimes been referred to as generalized breathers \cite{Groves}.
Over longer time windows, the amplitude of the signal can grow again
(leading to a repeated appearance of breathers), but  eventually the
structure typically breaks down, leading to chaotic type dynamics for long-time simulations. Similar observations
have been made for k-gap solitons in photonic systems \cite{Superluminal}.
The existence of perfectly homoclinic solutions in this system is an important question,
but it lies beyond the scope of the present article.
Inspection of the long time dynamics, as well as breathers for other
parameter values, are provided in the Supplemental Material. 
We now generate a family of wavenumber bandgap breathers parameterized
by the distance the underlying wavenumber is from the band edge $q_r$. 
This is a natural parameter to consider, as the distance to the bandedge
determines breather width and amplitude in space-periodic systems \cite{flach_discrete_2008}.
Keeping all parameters fixed, but gradually varying the modulation
frequency $f_{\rm mod}$ has the effect of shifting the bandgap in the wavenumber axis. Thus,
we fix the wavenumber ($q_5$ in this case), whose distance
to the right edge will increase as the modulation frequency is increased.
%(how exactly is quantified in the Supplemental Material).
For the breather shown in Fig.~\ref{fig:breather_ns10}
the distance to the edge is $\Delta q  = q_r - q_5 \approx 0.09$.
The red dots of Fig.~\ref{fig:cont}(a) show the amplitude
of the breather vs. $\Delta q  = q_r(f_{\rm mod}) - q_5$. 
%Note that
 %the right edge $q_r = q_r(f_{\rm mod})$ is now a function of the modulation
%frequency $f_{\rm mod}$. 
The amplitude is computed as the maximum
 velocity of the 6th node, i.e.,  $ \max_t \| v_6(t) \|$.
For $\Delta q > 0.4$ the breathers do not form a coherent localization, like in Fig.~\ref{fig:breather_ns10}.
Similar simulations with wavenumber near
the left edge of the bandgap ($q_5 \approx q_\ell$) did not lead to the robust formation
of breathers. 
%% This paragraph below could go.
The amplitude data is fit with a function of the form
$\gamma_1 \Delta q^{\gamma_2} $, with the best fit values being $\gamma_1=3.34$ and $\beta_2 = 0.57$
(the solid line in Fig.~\ref{fig:cont}(a)).
This is consistent with the trend found for classic breathers in space-periodic systems
where it is well known that the breather amplitude grows like $\mathcal{O}\left(\sqrt{\Delta \omega}\right)$,
where $\Delta \omega$ is the difference between the breather frequency 
and the edge of the frequency spectrum \cite{flach_discrete_2008}. 

Now that we have established the existence of wavenumber bandgap breathers
in Eq.~\eqref{eq:EOM}, we now consider the role of damping, which will bring
us closer to the experimentally relevant situation.
Breathers in experiments will always be a dissipative analog of breathers in lossless models.
Thus, we modify our definition of a breather to 
account for dissipative effects. To motivate our modified definition,
we repeat the simulations shown in Fig.~\ref{fig:breather_ns10} but
with a nonzero damping parameter. The yellow square (blue circle) markers of Fig.~\ref{fig:breather_ns10}(c)
show the orbit with a damping parameter of $c=0.075$ ($c=0.15$) Ns/m. The orbit starts close
to being homoclinic, but the dynamics are attracted to a stable fixed point
(i.e., a time-periodic orbit of the original system with period $2T$).
Note that the time-dependent stiffness term in the model (with the $k(t)$ coefficient) acts
as an effective gain term to balance loss.
%This fixed point exists in the lossless ($c=0$) system too, but is neutrally stable.
The orbit in the damped system experiences an initial exponential growth
and a turning point, like the lossless breather, but rather than approaching
near zero amplitude, the dynamics tend to a stable fixed point. 
Thus, the
left tail of the ``damped breather" (in the time domain) is much like a lossless breather,
whose amplitude is slightly lower due to the presence of damping. The right
tail of the ``damped breather" approaches a periodic oscillation, whose amplitude is
not necessarily small relative to the amplitude of the breather.
See Fig.~\ref{fig:cont}(b) for an example time series with
$c = 0.15$ Ns/m.  An alternative
classification for the structure found in the damped system would be a
wavenumber bandgap ``front", since the solution is heteroclinic, 
as it connects the unstable zero fixed point to a non-zero stable fixed point.
Important for the present study, however, is that the transition between the zero and nonzero state is approximately described 
by the lossless breather (i.e., the near homoclinic orbit).
We
measure the amplitude of the structure in the same way we measured
the breather amplitude in the lossless system, i.e.,  $ \max_t \| v_6(t) \|$.
We repeat this for various modulation frequencies (and hence $\Delta q$ values)
for the damping value $c=0.15$ Ns/, which is shown
as the blue circle markers in Fig.~\ref{fig:cont}(a).
The qualitative amplitude trend is similar to the lossless case, but
the amplitude is decreased.
The amplitude data in the damped case is also fit with a function of the form
$\gamma_1 \Delta q^{\gamma_2} $, with the best fit values
being $\gamma_1=2.62$ and $\beta_2 = 0.59$. 

We now turn to the experimental construction of wavenumber bandgap breathers (using the dissipative definition
of a breather defined above). To excite a plane wave with a particular wavenumber (in this case $q_5$)
 the unmodulated system is driven with the frequency $\omega_{\rm disp}(q_5)$.
Once the desired plane wave is excited, the initial driving is turned off and the modulation is turned on simultaneously. Like
in the damped simulation, the amplitude will initially grow, reach a turning point,
decay, but eventually approach a periodic orbit (or exhibit chaotic behavior).
An example experimental time series is shown in Fig.~\ref{fig:cont}(c).
Notice the qualitative agreement to the theoretical prediction shown in Fig.~\ref{fig:cont}(b).
Since the band edge can only be computed in the infinite lattice,
we use the relation $\Delta q  = q_r(f_{\rm mod}) - q_5$, where $q_r(f_{\rm mod})$
is found analytically. In general, there will be a mistuning between the experimental and model dynamics
for a fixed modulation frequency. We account for this mistuning
by measuring the modulation frequency the mode $q_5$ becomes unstable in the model
and experiment. The difference between the experimental
and theoretical critical modulation frequencies, $\delta f_{\rm mod}$,
is then added to the experimental modulation frequencies when calculating
the distance to the band edge, namely $\Delta q  = q_r(f_{\rm mod} - \delta f_{\rm mod}) - q_5$.
Thus, $\Delta q =0$ corresponds to the frequency at which $q_5$ becomes unstable
for both experiment and model. The  mode associated to $q_5$ is considered unstable when
its corresponding amplitude in Fourier space exceeds a noise threshold
(details given in the Supplemental Material).
With these definitions in place, we now
measure the amplitude of the breather as a function of $\Delta q$,
see the markers with error bars in Fig.~\ref{fig:cont}(a).
The qualitative amplitude trend agrees with the theoretical prediction.
Although the time series between the
theoretical prediction for $c=0.15$ Ns/m and the experiment 
agree qualitatively (compare panels (b) and (c) of Fig.~\ref{fig:cont}) the model
underestimates the amplitude. This can be partially explained
by the fact that other modes (including unstable ones) in the experiment besides $q_5$ are excited. Inspection of the
spatial Fourier transform of the experiment shows
there are always traces of additional modes, even before
the turning point of the structure. The insets of 
Fig.~\ref{fig:cont}(c) show the spatial Fourier
transform before ($t=1.6$ s) and after ($t=2.2$ s) the turning point
(more details of the Fourier analysis of the experiment
is given in the Supplemental Material).

Finally, we measure the width of the structures using the half-width at half maximum (HWHM) metric.
The HWHM is given by $t_{\rm max} - t_{\rm half}$, where $t_{\rm max}$ is the time the maximum is attained
and $t_{\rm half}$ is the time where the trajectory first attains half the maximum value.
The experimentally measured values are shown as the black triangles in Fig.~\ref{fig:cont}(a).
The width of the structure can be predicted theoretically using the real part of the Floquet exponent.
In particular, assuming the envelope of the breather follows the form $a\sech( \sigma t )$ (which we demonstrated
previously was a reasonable assumption), we can compute the HWHM as  $\sech^{-1}(1/2) / \sigma_5$.
The prediction in the lossless ($c=0$) and damped ($c=0.15$ Ns/m) cases are shown
as red solid squares and open blue squares, respectively, in Fig.~\ref{fig:cont}(a).
For sufficiently large values of $\Delta q$, there is good agreement between theory
and experiment, and both show the structure becomes more narrow as the wavenumber
goes deeper into the gap. This is consistent with classic breathers
in space-periodic systems.
\vspace{-.5cm}

 \section{Conclusion}
 \vspace{-.25cm}
 We explored a new class of solutions, 
phononic wavenumber bandgap breathers and their damped analogues,
in a time-varying lattice.
 Several avenues of inquiry follow naturally from this study, including the possible existence 
 of genuine q-gap breathers (i.e., with both tails decaying to zero),
 an analytical description of the breather profiles via a multiple-scale analysis,
 and the exploration of such structures in higher spatial dimensions.

\emph{Acknowledgement}.
This material is based upon work supported by the US National Science Foundation under Grant Nos.
EFRI-1741565 (C.D.) and DMS-2107945 (C.C. and E.W.). The authors are grateful for discussions with P.G.~Kevrekidis.

\bibliographystyle{unsrt}
%\bibliography{../../../../Chong}
\bibliography{LatticeManuscript2}

\end{document}